\def\eck#1{\left\lbrack #1 \right\rbrack}

\def\rund#1{\left( #1 \right)}
\def\abs#1{\left\vert #1 \right\vert}

\def\ave#1{\left\langle #1 \right\rangle}

\def\part#1#2{{\partial #1\over\partial #2}}

\def\A{{\cal A}}

\def\L{{\cal L}}

\def\d{{\rm d}}

\def\arcsecf {\hbox{$.\!\!^{\prime\prime}$}}

\def\vp{\varphi}

\def\Real{{\rm I\mathchoice{\kern-0.70mm}{\kern-0.70mm}{\kern-0.65mm}%
  {\kern-0.50mm}R}}
\def\C{\rm C\kern-.42em\vrule width.03em height.58em depth-.02em
       \kern.4em}
\font \bolditalics = cmmib10
\def\bx#1{\leavevmode\thinspace\hbox{\vrule\vtop{\vbox{\hrule\kern1pt
        \hbox{\vphantom{\tt/}\thinspace{\bf#1}\thinspace}}
      \kern1pt\hrule}\vrule}\thinspace}

\def \vc #1{{\textfont1=\bolditalics \hbox{$\bf#1$}}}
{\catcode`\@=11
\gdef\SchlangeUnter#1#2{\lower2pt\vbox{\baselineskip 0pt \lineskip0pt
  \ialign{$\m@th#1\hfil##\hfil$\crcr#2\crcr\sim\crcr}}}
}

\def\lesssim{\mathrel{\mathpalette\SchlangeUnter<}}

\def\ueber#1#2{{\setbox0=\hbox{$#1$}%
  \setbox1=\hbox to\wd0{\hss$\scriptscriptstyle #2$\hss}%
  \offinterlineskip
  \vbox{\box1\kern0.4mm\box0}}{}}

\def\bx#1{\leavevmode\thinspace\hbox{\vrule\vtop{\vbox{\hrule\kern1pt
        \hbox{\vphantom{\tt/}\thinspace{\bf#1}\thinspace}}
      \kern1pt\hrule}\vrule}\thinspace}

\def\SFB{{This work was supported by the ``Sonderforschungsbereich
375-95 f\"ur
Astro--Teil\-chen\-phy\-sik" der Deutschen For\-schungs\-ge\-mein\-schaft.}}
 
\magnification=\magstep1
\input epsf
\voffset= 0.0 true cm
\vsize=19.8 cm     
\hsize=13.5 cm
\hfuzz=2pt
\tolerance=500
\abovedisplayskip=3 mm plus6pt minus 4pt
\belowdisplayskip=3 mm plus6pt minus 4pt
\abovedisplayshortskip=0mm plus6pt
\belowdisplayshortskip=2 mm plus4pt minus 4pt
\predisplaypenalty=0
\footline={\tenrm\ifodd\pageno\hfil\folio\else\folio\hfil\fi}

\def\la{\mathrel{\hbox{\rlap{\hbox{\lower4pt\hbox{$\sim$}}}\hbox{$<$}}}}
\def\ga{\mathrel{\hbox{\rlap{\hbox{\lower4pt\hbox{$\sim$}}}\hbox{$>$}}}}

\def\arcsec{\hbox{$^{\prime\prime}$}}
\def\utw{\smash{\rlap{\lower5pt\hbox{$\sim$}}}}
\def\udtw{\smash{\rlap{\lower6pt\hbox{$\approx$}}}}

\def\getsto{\mathrel{\hbox{\rlap{$\gets$}\hbox{\raise2pt\hbox{$\to$}}}}}
\def\lid{\mathrel{\hbox{\rlap{\hbox{\lower4pt\hbox{$=$}}}\hbox{$<$}}}}
\def\gid{\mathrel{\hbox{\rlap{\hbox{\lower4pt\hbox{$=$}}}\hbox{$>$}}}}
\def\sol{\mathrel{\hbox{\rlap{\hbox{\raise4pt\hbox{$\sim$}}}\hbox{$<$}}}
}
\def\sog{\mathrel{\hbox{\rlap{\hbox{\raise4pt\hbox{$\sim$}}}\hbox{$>$}}}
}
\def\lse{\mathrel{\hbox{\rlap{\hbox{\raise4pt\hbox{$<$}}}\hbox{$\simeq$}
}}}
\def\gse{\mathrel{\hbox{\rlap{\hbox{\raise4pt\hbox{$>$}}}\hbox{$\simeq$}
}}}
\def\grole{\mathrel{\hbox{\lower2pt\hbox{$<$}}\kern-8pt
\hbox{\raise2pt\hbox{$>$}}}}
\def\leogr{\mathrel{\hbox{\lower2pt\hbox{$>$}}\kern-8pt
\hbox{\raise2pt\hbox{$<$}}}}
\def\loa{\mathrel{\hbox{\rlap{\hbox{\lower4pt\hbox{$\approx$}}}\hbox{$<$
}}}}
\def\goa{\mathrel{\hbox{\rlap{\hbox{\lower4pt\hbox{$\approx$}}}\hbox{$>$
}}}}

%
%

\font\kleinhalbcurs=cmmib10 scaled 833
\font\eightrm=cmr8
\font\sixrm=cmr6
\font\eighti=cmmi8
\font\sixi=cmmi6
\skewchar\eighti='177 \skewchar\sixi='177
\font\eightsy=cmsy8
\font\sixsy=cmsy6
\skewchar\eightsy='60 \skewchar\sixsy='60
\font\eightbf=cmbx8
\font\sixbf=cmbx6
\font\eighttt=cmtt8
\hyphenchar\eighttt=-1
\font\eightsl=cmsl8
\font\eightit=cmti8

\font\bxf=cmbx10
  \mathchardef\Gamma="0100
  \mathchardef\Delta="0101
  \mathchardef\Theta="0102
  \mathchardef\Lambda="0103
  \mathchardef\Xi="0104
  \mathchardef\Pi="0105
  \mathchardef\Sigma="0106
  \mathchardef\Upsilon="0107
  \mathchardef\Phi="0108
  \mathchardef\Psi="0109
  \mathchardef\Omega="010A
\def\rahmen#1{\vskip#1truecm}
\def\begfig#1cm#2\endfig{\par
\setbox1=\vbox{\rahmen{#1}#2}%
\dimen0=\ht1\advance\dimen0by\dp1\advance\dimen0by5\baselineskip
\advance\dimen0by0.4true cm
\ifdim\dimen0>\vsize\pageinsert\box1\vfill\endinsert
\else
\dimen0=\pagetotal\ifdim\dimen0<\pagegoal
\advance\dimen0by\ht1\advance\dimen0by\dp1\advance\dimen0by1.4true cm
\ifdim\dimen0>\vsize
\topinsert\box1\endinsert
\else\vskip1true cm\box1\vskip4true mm\fi
\else\vskip1true cm\box1\vskip4true mm\fi\fi}
\def\figure#1#2{\smallskip\setbox0=\vbox{\noindent\petit{\bf Fig.\ts#1.\
}\ignorespaces #2\smallskip
\count255=0\global\advance\count255by\prevgraf}%
\ifnum\count255>1\box0\else
\centerline{\petit{\bf Fig.\ts#1.\ }\ignorespaces#2}\smallskip\fi}

\def\xfigure#1#2#3#4{\midinsert\noindent
    $$\epsfxsize=#4truecm\epsffile{#3}$$
    \figure{#1}{#2}\endinsert}


\def\begtab#1cm#2\endtab{\par
\ifvoid\topins\midinsert\vbox{#2\rahmen{#1}}\endinsert
\else\topinsert\vbox{#2\kern#1true cm}\endinsert\fi}
\def\rahmen#1{\vskip#1truecm}
\def\begpet{\vskip6pt\bgroup\petit}
\def\endpet{\vskip6pt\egroup}
\def\begref{\par\bgroup\petit
\let\it=\rm\let\bf=\rm\let\sl=\rm\let\INS=N}
\def\petit{\def\rm{\fam0\eightrm}%
\textfont0=\eightrm \scriptfont0=\sixrm \scriptscriptfont0=\fiverm
 \textfont1=\eighti \scriptfont1=\sixi \scriptscriptfont1=\fivei
 \textfont2=\eightsy \scriptfont2=\sixsy \scriptscriptfont2=\fivesy
 \def\it{\fam\itfam\eightit}%
 \textfont\itfam=\eightit
 \def\sl{\fam\slfam\eightsl}%
 \textfont\slfam=\eightsl
 \def\bf{\fam\bffam\eightbf}%
 \textfont\bffam=\eightbf \scriptfont\bffam=\sixbf
 \scriptscriptfont\bffam=\fivebf
 \def\tt{\fam\ttfam\eighttt}%
 \textfont\ttfam=\eighttt
 \normalbaselineskip=9pt
 \setbox\strutbox=\hbox{\vrule height7pt depth2pt width0pt}%
 \normalbaselines\rm
\def\vec##1{\setbox0=\hbox{$##1$}\hbox{\hbox
to0pt{\copy0\hss}\kern0.45pt\box0}}}%
\let\ts=\thinspace
%
\font \tafontt=     cmbx10 scaled\magstep2
\font \tafonts=     cmbx7  scaled\magstep2
\font \tafontss=     cmbx5  scaled\magstep2
\font \tamt= cmmib10 scaled\magstep2
\font \tams= cmmib10 scaled\magstep1
\font \tamss= cmmib10
\font \tast= cmsy10 scaled\magstep2
\font \tass= cmsy7  scaled\magstep2
\font \tasss= cmsy5  scaled\magstep2
\font \tasyt= cmex10 scaled\magstep2
\font \tasys= cmex10 scaled\magstep1
\font \tbfontt=     cmbx10 scaled\magstep1
\font \tbfonts=     cmbx7  scaled\magstep1
\font \tbfontss=     cmbx5  scaled\magstep1
\font \tbst= cmsy10 scaled\magstep1
\font \tbss= cmsy7  scaled\magstep1
\font \tbsss= cmsy5  scaled\magstep1

\newbox\chsta\newbox\chstb\newbox\chstc
\def\centerpar#1{{\advance\hsize by-2\parindent
\rightskip=0pt plus 4em
\leftskip=0pt plus 4em
\parindent=0pt\setbox\chsta=\vbox{#1}%
\global\setbox\chstb=\vbox{\unvbox\chsta
\setbox\chstc=\lastbox
\line{\hfill\unhbox\chstc\unskip\unskip\unpenalty\hfill}}}%
\leftline{\kern\parindent\box\chstb}}
 \def \chap#1{
    \vskip24pt plus 6pt minus 4pt
    \bgroup
 \textfont0=\tafontt \scriptfont0=\tafonts \scriptscriptfont0=\tafontss
 \textfont1=\tamt \scriptfont1=\tams \scriptscriptfont1=\tamss
 \textfont2=\tast \scriptfont2=\tass \scriptscriptfont2=\tasss
 \textfont3=\tasyt \scriptfont3=\tasys \scriptscriptfont3=\tenex
     \baselineskip=18pt
     \lineskip=18pt
     \raggedright
     \pretolerance=10000
     \noindent
     \tafontt
     \ignorespaces#1\vskip7true mm plus6pt minus 4pt
     \egroup\noindent\ignorespaces}%
 \def \sec#1{
     \vskip15true pt plus4pt minus4pt
     \bgroup
 \textfont0=\tbfontt \scriptfont0=\tbfonts \scriptscriptfont0=\tbfontss
 \textfont1=\tams \scriptfont1=\tamss \scriptscriptfont1=\kleinhalbcurs
 \textfont2=\tbst \scriptfont2=\tbss \scriptscriptfont2=\tbsss
 \textfont3=\tasys \scriptfont3=\tenex \scriptscriptfont3=\tenex
     \baselineskip=16pt
     \lineskip=16pt
     \raggedright
     \pretolerance=10000
     \noindent
     \tbfontt
     \ignorespaces #1
     \vskip12true pt plus4pt minus4pt\egroup\noindent\ignorespaces}%
 \def \subs#1{
     \vskip8true pt plus 4pt minus4pt
     \bgroup
     \bxf
     \noindent
     \raggedright
     \pretolerance=10000
     \ignorespaces #1
     \vskip6true pt plus4pt minus4pt\egroup
     \noindent\ignorespaces}%
 \def \subsubs#1{
     \vskip15true pt plus 4pt minus 4pt
     \bgroup
     \bf
     \noindent
     \ignorespaces #1\unskip.\ \egroup
     \ignorespaces}
\def\footnoterule{\kern-3pt\hrule width 2true cm\kern2.6pt}
\newcount\footcount \footcount=0
\def\advftncnt{\advance\footcount by1\global\footcount=\footcount}
\def\fonote#1{\advftncnt$^{\the\footcount}$\begingroup\petit
       \def\textindent##1{\hang\noindent\hbox
       to\parindent{##1\hss}\ignorespaces}%
\vfootnote{$^{\the\footcount}$}{#1}\endgroup}

\newcount\sterne
\outer\def\byebye{\bigskip\typeset
\sterne=1\ifx\speciali\undefined\else
\bigskip Special caracters created by the author
\loop\smallskip\noindent special character No\number\sterne:
\csname special\romannumeral\sterne\endcsname
\advance\sterne by 1\global\sterne=\sterne
\ifnum\sterne<11\repeat\fi
\vfill\supereject\end}
\def\typeset{\centerline{\petit This article was processed by the author
using the \TeX\ Macropackage from Springer-Verlag.}}
 
\voffset=0pt


\def\fprime{f^\prime}
\def\bx{x}
\def\psig{\psi_{\rm g}}
\def\psis{\psi_\gamma}
\def\dpsi{\delta\psi}
\def\apj{ApJ}
\def\aj{AJ}
\def\mnras{MNRAS}

\def\obs{{({\rm obs})}}
\chap{Evidence for substructure in lens galaxies?}
\medskip
\centerline{\bf Shude Mao \& Peter Schneider}
\centerline{\bf Max-Planck-Institut f\"ur Astrophysik}
\centerline{\bf Postfach 1523}
\centerline{\bf D-85740 Garching, Germany}
\bigskip
\sec{Abstract}
We discuss whether one should expect that multiply imaged QSOs can be
understood with `simple' lens models which contain a handful of
parameters. Whereas for many lens systems such simple mass models
yield a remarkably good description of the observed properties, there
are some systems which are notoriously difficult to understand
quantitatively. We argue that at least in one case (B 1422+231) these
difficulties are not due to a `wrong' parametrization of the lens
model, but that the discrepancy between observed and model-predicted
flux ratios are due to substructure in the lens. Similar to
microlensing for optical fluxes, such substructure can distort also the
radio flux ratios predicted by `simple' mass models, in particular for
highly magnified images, without appreciably
changing image positions. Substructure also does not change
the time delay significantly and therefore has 
little effect on the determination of the Hubble constant 
using time delays. We quantify these statements with several simple
scenarios for substructure, and propose a strategy to model lens
systems in which substructure is suspected.

\sec{1 Introduction}
Multiply imaged QSOs and radio (Einstein) rings provide the most
accurate mass measures of distant galaxies (Zwicky 1937; for specific
examples, see e.g. Rix, Schneider \& Bahcall 1992; Wallington,
Kochanek \& Narayan 1996) and promise to
provide one of the most robust methods for measuring $H_0$ (Refsdal
1964; for recent examples, see Falco et al.\ts 1996; Courbin et al.\ts
1997).\fonote{For a recent overview on the determination of $H_0$ from
lenses, see the proceedings of the Jodrell Bank meeting on {\it Golden
Lenses}, held at Jodrell Bank on June 23--25, 1997, available at 
http://multivac.jb.man.ac.uk:8000/ceres/workshop1/proceedings.html}
These successes and expectations are based on our ability to
understand the lensing geometry in sufficient detail. It is truly 
remarkable that many multiple QSOs can in fact be modelled
quite accurately with a simple elliptical deflection potential or an
elliptical mass distribution. This has nourished the expectation that most
lens systems are in fact due to fairly simple mass distributions (a
well-known exception is the system MG2016+112; see Nair \& Garrett
1997).

When modelling a multiple QSO lens system, one can either include or
disregard the flux ratios of the images. Given that the number of
observational constraints in these systems is never much larger than
the number of free parameters of the lens model (and often is the
same), there is of course a strong motivation to make use of the flux
ratio information. As pointed out by Chang \& Refsdal (1979), this may
be a dangerous undertaking, given that the sizes of the optical
continuum emitting regions of QSOs are expected to be of the same order
as the Einstein radius of a star in the lens galaxy, so that the
optical magnitudes may well be affected by gravitational microlensing
(see Wambsganss 1990, and references therein), even if averaged over
long periods of time.  For example in the case of QSO2237+0305, the
image positions can be fitted very accurately with a variety of lens
models (e.g., Kent \& Falco 1988, Rix et al.\ 1992, Wambsganss \&
Paczy\'nski 1994), but the observed optical flux of image D is smaller than
predicted, whereas the radio flux ratios (Falco et al.\ 1996) are in
much better accord with the lens models. Therefore, constraints
derived from optical fluxes should be used with care only, whereas
radio sources are expected to be extended much beyond the Einstein
radius of a stellar mass object, so that their flux should be
(largely) unaffected by microlensing and thus provide useful
constraints for lens modelling.

That argument would of course be weakened if lens galaxies contain
mass clumps with an Einstein radius comparable to the size of the
radio components -- i.e., masses of the order of, or exceeding,
globular clusters. In that case the same situation as microlensing in
the optical would apply. But this `milli-lensing' could cause
observable image splittings (Wambsganss \& Paczy\'nski 1992)
which have not yet been detected, so that
this possibility appears not very plausible. 

Nevertheless, as we shall argue in Sect.\ts 2, the difficulty in modelling
some lens systems in detail, with radio flux ratios included, points
towards the possibility of
substructure in lens galaxies (or somewhere else along the
line-of-sight to the QSO). Two kinds of substructure are briefly
considered in Sect.\ts 3, and numerical experiments to determine the
probability of appreciable flux-ratio changes are presented in
Sect.\ts 4. In Sect.\ts 5, we offer a practical way of
treating (radio) flux ratios in the presence of substructure.
We summarize our results in Sect.\ts 6.

\sec{2 B1422+231: More than a challenge for lens modellers}
The quadruply imaged QSO 1422+231 at $z_s=3.62$
was discovered in the course of the
JVAS survey (Patnaik et al.\ 1992). The four images have a maximum
separation of $1\arcsecf3$, and the lens galaxy has been accurately
located (Impey et al.\ 1996) and its redshift ($z_{\rm d}=0.34$) has
recently been measured
(Kundic et al.\ts 1997, Tonry 1997).
The flux ratios of the images
are different in the radio and optical bands; in the radio, they are
A:B:C:D$=$ 0.98:1:0.52:0.02, whereas in the optical, image A is
fainter than B, yielding A:B$\approx$0.8:1, somewhat dependent on
the optical filter (the radio flux ratios are nearly independent of
radio frequency), whereas the B:C:D ratios are largely compatible with
the radio. Given that the optical flux may be affected by microlensing
and/or dust obscuration, the radio flux ratios should be used in
modelling this system.

Several serious attempts have been made to model the lens in this
system (e.g., Hogg \& Blandford 1993, Kormann, Schneider \& Bartelmann
1994b, Keeton, Kochanek \& Seljak 1997) -- and they all failed!
Whereas the image positions can be fitted very accurately, the radio
flux ratios could not be obtained. Given that the parametrized lens
models used by the different authors differ moderately, this
failure is probably not due to a too restricted choice of the families
of lens models. Rather it may be a generic difficulty for `simple' (i.e.,
smooth) lens models, as can be seen as follows.

The large flux ratio between each of the images A, B, C, and the image D
$\sim 50$, leaves two possibilities: (i) image D is highly
demagnified, or (ii) A, B, C are highly magnified. We can exclude
option (i), since the separation of D from the center of the lens
galaxy -- $0\arcsecf 3$ -- does not put it into a region of very high
surface mass density as is necessary to obtain a strong
demagnification. This leaves option (ii) only.
Three highly magnified images occur generically when a
source is close to, and inside a cusp (see Chap.\ts 6 of Schneider,
Ehlers \& Falco 1992). In that case, there exists a universal relation
between the image fluxes (see, e.g., Schneider \& Weiss
1992; Mao 1992), namely that the sum of the fluxes of the outer two images (A
and C) should equal the flux of the middle image (B). This relation is
grossly violated in 1422+231, although the VLBI structure (Patnaik \&
Porcas 1997) shows a strong
elongation in the direction tangent to the direction to the lens
galaxy thus supporting the cusp hypothesis.  Hence, although this
universal relation is strictly valid only asymptotically for very high
magnification, or in other words, if the source is sufficiently close
to the cusp (and the theory of the lens mapping near cusps
developed so far does not provide a good handle of where this
`asymptotic regime begins'), this argument easily explains why all the
published lens models yield flux ratios with (A$+$C)/B$\lesssim$1.3 (the
observed value in the radio is 1.5). A
strong violation of the cusp constraint can be obtained either by 
putting the source further away from the cusp -- that 
would decrease the magnification, thus the magnification ratio to D,
therefore running into the same problem as option (i) above -- or
to impose a small-scale structure on the lens model which can locally
change the magnification of individual images. Here, small scale means
a scale smaller than the separation between A, B, and C, i.e., smaller
than $\sim 0\arcsecf 5$.

Although B1422+231 is the system where problems in the modelling are
most easily seen, it may not be unique. Keeton et al.\ (1997) and
Falco, Leh\'ar \& Shapiro (1996) have modelled MG0414+0534, and in
both attempts the resulting $\chi^2$ per degree of freedom is too
large to view these models as satisfying. In this system, the
components A1 and A2 have a separation much smaller than to the
distance to the other
two components. This suggests that they lie close to a critical curve
passing between them. In that case, one would expect the fluxes
of A1 and A2 has to be much larger than those of B and C, and that the flux
ratio A1/A2 is of order unity. This expectation is not at all
satisfied with the optical data, and so again one should not be
surprised to find that `simple' models fail (for this case, microlensing
by stellar mass objects offers a plausible explanation, see
Witt, Mao \& Schechter 1995). A similar situation occurs
in PG1115+080 (Keeton \& Kochanek 1996) where all the $\chi^2$ of the
best fits comes from the flux ratio of A1/A2, which is significantly
different from unity (Courbin et al.\ts 1997), even though a much
larger flux uncertainty is assumed for the individual
images than the measurement uncertainty in order to account for
microlensing. We consider these
difficulties as a hint that substructure may be present
in the lens galaxies. In the
next two sections, we shall show how small amplitude `perturbations' of
the mass distribution in the lens are sufficient to lead to the
observed discrepancies. 

\sec{3 Analytic estimates}
We shall consider two particular hypotheses for the kind of
substructure that may be present in a lens galaxy like that in B1422+231:
globular clusters (or in lensing language: point masses of order $10^6
M_\odot$), and small amplitude fluctuations.

\subs{3.1 The `globular cluster' picture}
A point mass of $\sim 10^6 M_\odot$ yields a deflection angle at its
Einstein radius of about 1 milli-arcsecond, and thus can cause multiple
images with similar separation, provided that source and lens are
sufficiently well aligned. Since we are interested in small
perturbations of the magnification of a macroimage, the typical
situation will be one where the separation of the point mass from the
image is considerably larger than the Einstein radius, so that the
deflection angle is also smaller, and the secondary image will be
highly demagnified. 

Consider a macroimage with local surface mass density $\kappa_0$ and
shear $\gamma_0$, with coordinates chosen such that only the 1-component
of the shear caused by the macro-model is non-zero. Let a point
mass be located at position $(x \cos\vp,x\sin\vp)$ relative to the image,
where $x$ is measured in units of the Einstein radius of the point
mass. The unperturbed and perturbed magnifications are then,
respectively, 
$$
\mu_0={1\over (1-\kappa_0)^2-\gamma_0^2}\quad ;\quad
\mu={1\over
(1-\kappa_0)^2-(\gamma_0+\delta\gamma_1)^2-\delta\gamma_2^2}\;,
\eqno (1)
$$
where $(\delta\gamma_1,\delta\gamma_2)=(\cos 2\vp,\sin 2\vp)/x^2$ is
the shear caused by the point mass lens. Let $R=\mu_0/\mu$ be the
ratio of the unperturbed to the perturbed magnification; its
dependence on $x$ and $\vp$ can be written as
$$
x^2={\gamma_0 \mu_0\cos2\vp\over 1-R}\rund{1+\sqrt{1+{1-R\over \mu_0 
(\gamma_0 \cos2\vp)^2}}}
\approx {2 \gamma_0 \mu_0\cos2\vp\over 1-R}\;,
\eqno (2)
$$
where in the second step we assumed that $\abs{1-R}\ll 1$, since we are
interested in small magnification perturbations. Thus, locations $\vc
x$ of the point mass for constant flux perturbation $R$ trace an
`$\infty$'-like curve. The area of the region in which a point mass
has to be located in order to cause a flux perturbation larger
(smaller) than $R$ for $R>1$ ($R<1$) is
$$
A=2\abs{ \gamma_0 {\mu_0\over 1-R}}\; ,
\eqno (3)
$$ 
as can be easily obtained from integrating (2) over $\vp$. 
For the lens models applicable to B1422+231, the A image should have
$\mu_0\sim 15$, $\gamma_0\sim0.5$, and so the area in which a point
mass lens has to be located to cause a flux perturbation of $\pm 20\%$
is $A\sim 150$. Thus, with a surface mass density $\kappa_*\sim 0.01$,
one obtains a high probability to find one of these point-mass lenses
inside the region where it causes an appreciable magnification change.

\subs{3.2 Plane wave perturbations}
As a second `model' for lens perturbations, we consider a plane
density wave. For simplicity of the argument, we again assume that the
shear caused by the macromodel has only a 1-component, and that the
plane wave is modulated also in the 1-direction. Then,
$\delta\kappa=\delta\gamma_1$, $\delta\gamma_2=0$. The ratio $R$ then
becomes, up to first order in the perturbation $\delta\kappa$,
$$
R=1-\delta\kappa{2\over 1-\kappa_0-\gamma_0}\;.
\eqno (4)
$$
The denominator in (4) is one of the eigenvalues of the magnification
matrix caused by the macromodel. For images near a critical curve, one
of the eigenvalues becomes very small, so that $R$ can deviate
significantly from unity even for small $\delta \kappa$. For the A
image of B1422+231, one expects $\kappa_0+\gamma_0 \sim 0.9$, so that for a
20\% change in magnification one needs $\delta\kappa\sim 0.01$, a
truly small fluctuation.

\sec{4. Numerical simulations}
We model the primary lensing galaxy G1 as 
a singular isothermal ellipsoid (SIE) density
distribution and take into account the effect of several nearby galaxies
as an external shear (Kormann, Schneider \& Bartelmann 1994b). The 
external shear may also include contributions from large scale structure
and/or the deviation of the primary lens galaxy from the SIE model.
We choose the coordinate system on the lens plane, $\vc{x}=(x_1, x_2)$,
centered on the primary lens galaxy, with the
$x_1$-axis along (decreasing) right ascension and the $x_2$-axis
along (increasing) declination (cf. Fig.\ts 3). The corresponding positions
in the source plane $\vc{y}=(y_1, y_2)$ have the
same orientations. The dimensionless coordinates
$\vc{x}$ and $\vc{y}$ are the observed angles
normalized to $\theta_0$, an 
angular scale (Einstein radius) which characterizes
the strength of the lens potential. We express
the surface density in units of the critical surface density,
$\Sigma_{\rm cr}$. For
an Einstein-De Sitter universe, $1\arcsec=3.0h^{-1} {\rm kpc}$ at the lens
plane, and the critical
surface density is $\Sigma_{\rm cr}=3650h M_\odot/{\rm pc}^2$, here $h$
is the present-day Hubble constant in units of 
$100~{\rm km~ s^{-1}~ Mpc^{-1}}$ and we take $h=0.5$. 

If the minor axis of the lens
is along the $x_1$-axis, the SIE potential can be written as
$$
\psig(x, \phi) = x {\sqrt{f} \over \fprime} \times 
[\sin\phi~\arcsin(\fprime \sin\phi)
+\cos\phi~{\rm arcsinh}~({\fprime \over f}\cos\phi)]\;,
\eqno (5)
$$
where $f$ is the axis ratio, $\fprime=\sqrt{1-f^2}$, and we have used
polar coordinates $\vc{x} = (x\cos\phi, x\sin\phi)$.
The shear perturbation is given by
$$
\psis(x, \phi) = {x^2\over 2} \gamma \cos(2\phi)\;,
\eqno (6)
$$
if the shear direction is along the $x_1$-axis. The total potential
is the sum of the SIE potential and the shear perturbation, 
$$
\psi_0=\psig(x, \phi-\phi_{\rm g}) + \psis(x, \phi-\phi_{\rm s})\;,
\eqno (7)
$$
where we have now generalized to the case when the lens minor axis 
encloses an angle $\phi_{\rm g}$, and the direction of shear
an angle of $\phi_{\rm s}$ with the $x_1$-axis.
The lens equation is then simply: $ \vc{y}=\vc{x}-\nabla \psi(\vc{x})$.
The Jacobian of the lens mapping is ${\cal A}=\partial \vc{y}/\partial \vc{x}$,
and the magnification of
an image is given by $\mu = ({\rm det} {\cal A})^{-1}$. We refer
the readers to Kormann et al. (1994a,b) for more technical details.

\subs{4.1 A macromodel}

Observational data is taken from Tables 1 and 2 in
Impey et al. (1996). The total number of observational constraints
is 12, namely we have $4\times 2$ image positions and 4
fluxes. Our lens model is described
by 8 parameters, i.e., three lens parameters, $(\phi_{\rm g}, f, \theta_0)$,
two shear parameters, $(\gamma,\phi_{\rm s})$, 
two source coordinates, and the unlensed source flux
$S_0$. Thus, the number of degrees of freedom
is 4. Nearly identical procedures to those in Kormann et al. (1994b)
are used to find the best lens model. A 
minor difference is that we have only included the flux ratio
D:B in the $\chi^2$ measure, since we argue that
substructures may have modified the flux ratios of the brighter
images significantly.
There are a few models that can fit the data with acceptable $\chi^2$
(cf. Keeton et al. 1997). One of these
is summarized in Table 1. The model
reproduces the observed positions within $1\sigma$ of the observational
uncertainties. The relative
magnifications are, however, unsatisfactory. The flux ratio 
A:B is 0.78, close to the observed optical flux ratio, 
but deviates significantly from the observed radio value 0.98.
In addition, image D is too bright compared with the observed value.

\vskip 0.5cm
\centerline{Table 1: Model Parameters}
\vskip 0.1cm
\vbox{
$$\vbox{ \halign{\hfil #\hfil && \quad \hfil #\hfil \hfil \hfil \hfil \hfil
\hfil\cr
\noalign{\hrule height0.6pt\vskip 2pt\hrule height0.6pt\vskip 1em}
$\theta_0 $ & $f$ & $\phi_g$ & $\gamma$ & $\phi_s $ 
& $\mu_A$ & $\mu_B$ & $\mu_C$ & $\mu_D$ \cr
\noalign{\vskip 0.5em\hrule height0.6pt\vskip 0.5em}
0.77 & 0.78 & $-1.01$ & 0.20 & $-1.04$ & 7.01 & $-9.01$ & 4.52 & $-0.33$ \cr
}}$$
}

\subs{4.2 The `globular cluster' picture}

The presence of massive lenses with mass $\sim 10^6 M_\odot$
provides a perturbation potential:
$$
\delta\psi = \hat{m} \sum_i
{1 \over 2} \ln [(x-x_i)^2+(y-y_i)^2], ~~
\hat{m} = {M \over \Sigma_{\rm cr} \pi (D_l \theta_0)^2}
\eqno(8)
$$
on top of $\psi_0$, where $\theta_0$ is again the Einstein radius,
$D_l=610h^{-1}$Mpc is the distance to the lens galaxy, and 
$M=5\times 10^5 M_\odot$ is the mass of the lenses.
For simplicity we assume that these lenses are uniformly 
distributed and have a surface density
of $\kappa_*$ in units of the critical surface density.
Similar to microlensing by
solar mass objects (Wambsganss 1990), we generate
point lenses randomly on the lens plane with the required
surface density.
To quantify the deviation of magnifications from that of the
unperturbed potential, we define a quantity
$r=(\mu_A+\mu_B+\mu_C)/(|\mu_A|+|\mu_B|+|\mu_C|)$. If the magnifications
of image A, B and C strictly obey the prediction of a cusp singularity,
$r=0$. The input value resulting from the unperturbed potential $\psi_0$ 
is $r=0.123$. In Fig.\ts 1, we show the probability distribution of
$r$ as obtained from Monte Carlo simulations. For $\kappa_*=0.005, 0.01$, the probability
of having a deviation
larger than the observed one is about 21\% and 36\%. Therefore,
a surface density in `globular clusters'
$\kappa_*\approx 0.005$ may be sufficient to explain the observed
deviation in flux ratios.
Nearly in all the cases, the change in the
image positions are a few milli-arcsecond. Such small changes are
undetectable in the optical. In any case, such small changes
can probably be accommodated by a small adjustment in the macromodel.

\xfigure{1}{
Probability distribution for the quantity
$r=(\mu_A+\mu_B+\mu_C)/(|\mu_A|+|\mu_B|+|\mu_C|)$
for a singular isothermal ellipsoid with external shear and massive
point lenses.
$r$ measures
the deviation of magnifications from the prediction ($r=0$) of
a cusp singularity.
Equal-mass point lenses are distributed uniformly with surface density
$\kappa_*$.
The dashed vertical line indicates the
input value for the unperturbed lens model.
The solid vertical line shows
the observed value of $r=0.2$.
Three curves are shown for $\kappa_*=0.0025, 0.005$ and $0.01$. 
The probability
of having a deviation larger
than the observed one is 11 (21, 36)\% for $\kappa_*=0.0025$ (0.005,
0.01), respectively.
}
{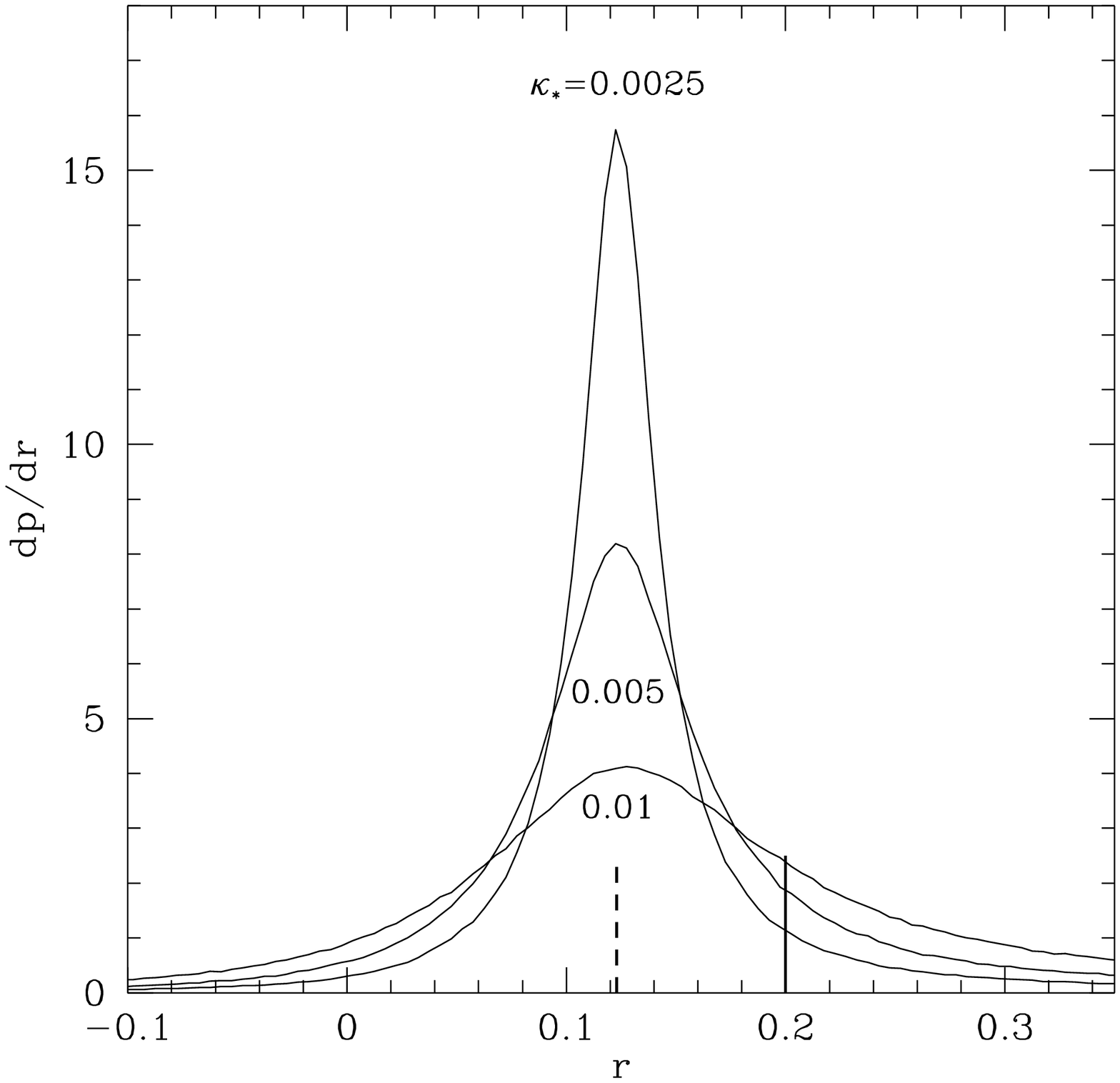}{10}

\subs{4.3 Plane wave perturbation}

For each image, we consider a plane-wave fluctuation centered on
the image, described by
$$
\delta \psi
 = {2 \kappa_s \over \abs{\vc k}^2}
 \cos(k_1 \theta_0 x_1 + k_2 \theta_0 x_2 + \phi_0) 
e^{-{x^2\theta^2_0 \over 2 \sigma_r^2}}\;,
\eqno (9)
$$
where $\kappa_s$ is the amplitude of fluctuation, $(k_1, k_2)$ is the
wave vector, $\phi_0$ is a phase constant, and the last exponential term
localises the perturbation within a few $\sigma_r$. Notice that
in the absence of the last term, the amplitude of surface density corresponding
to $\delta\psi$ is $\kappa_s$. Since the parameter space is rather
large, we limit our study to one illustrative case. We fix the 
magnitude of the wave-vector to be $2\pi/|\vc k|=0\arcsecf 01$ but with 
random orientations. $\phi_0$ is distributed uniformly between 0 to $2\pi$,
and the amplitude of fluctuations is drawn from a Gaussian
distribution
with zero mean and
standard deviation $\sigma_\kappa$. We take $\sigma_r=0\arcsecf 03$,
but the results are insensitive to the choice of $\sigma_r$.
In Fig.\ts 2, we show 
the probability distribution of $r$ for $\sigma_\kappa=0.005, 0.01, 0.02$. 
The probability
of having deviations larger
than the observed one is 1.5 (6.8, 13)\% for $\sigma=0.005$ (0.01, 
0.02), respectively. These values are in good agreement with
the analytical estimates presented in Sect.\ts 3.

\xfigure{2}{
Probability distribution for the quantity
$r=(\mu_A+\mu_B+\mu_C)/(|\mu_A|+|\mu_B|+|\mu_C|)$
for a singular isothermal ellipsoid with external shear and 
localised plane-wave fluctuations (cf. eq. [10]).
The amplitude of fluctuations
is drawn
from a Gaussian distribution with standard deviation $\sigma_\kappa$. Three
curves are shown with $\sigma_\kappa=0.005, 0.01, 0.02$.
The probability
of having deviations larger
than the observed one is 1.5 (6.8, 13)\% for $\sigma_\kappa=0.005$ (0.01,
0.02), respectively.
The dashed and solid
lines have the same meaning as those in Fig.\ts 1.
}{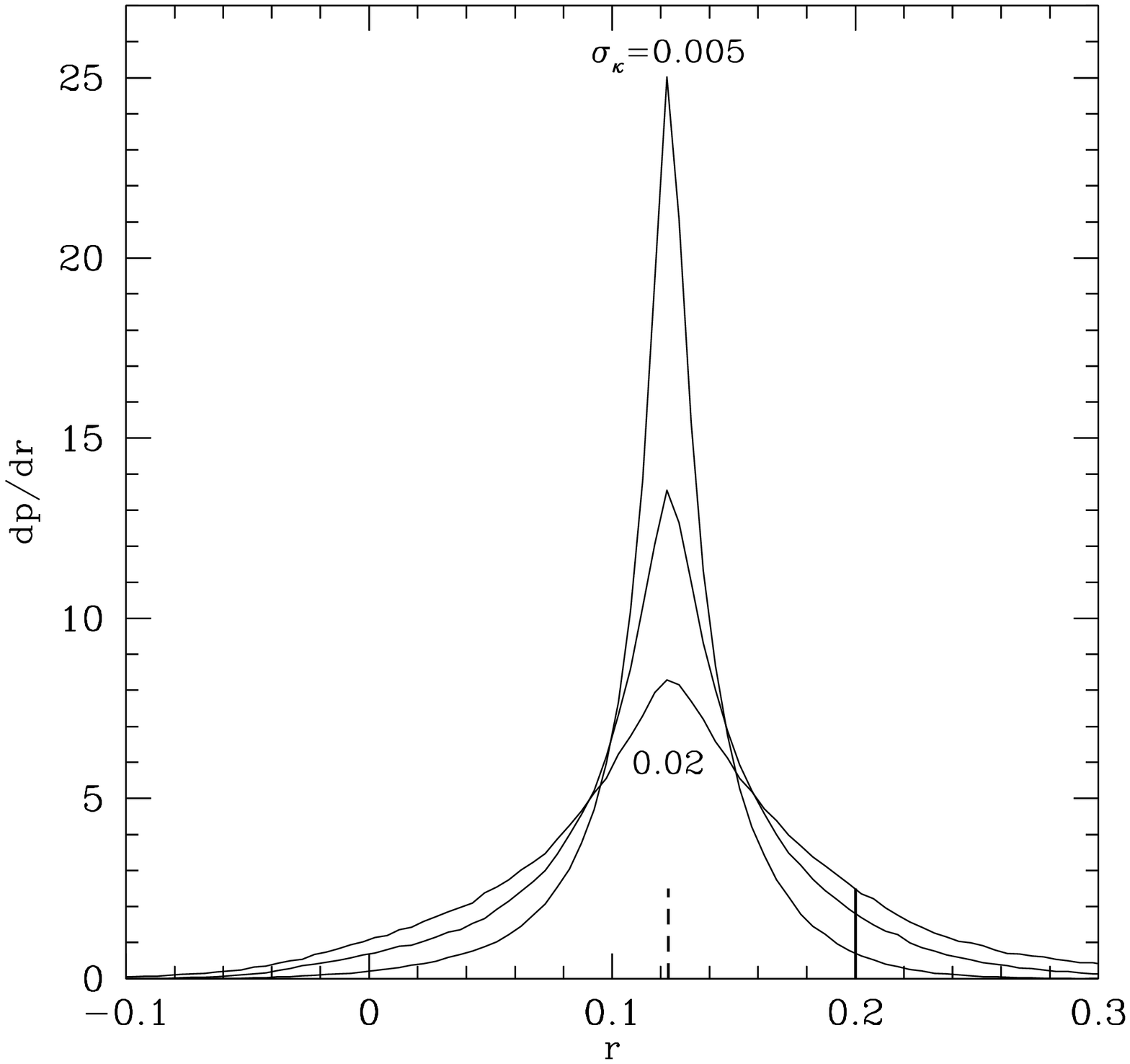}{10}

\subs{4.4 Size of fluctuation}

Let us estimate how small (in size) the smooth fluctuations have to
be in order to explain the observed deviation in B1422+231. We shall
consider a similar perturbation potential as that in Eq. (9), but without
the localisation exponential term.
For simplicity, we shall consider Fourier modes that are
stationary in a square with a side length, $L$, 
centered on the primary lens galaxy, i.e., the perturbation potential is
$$
\dpsi = \sum_{i,j=1}^N \sum_{l,m=0}^1
{2 \kappa_{ijlm} \over k^2_{ij}}
\cos(i {\pi \over L} \theta_0 x_1- {l \pi \over 2})
\cos(j {\pi \over L} \theta_0 x_2 -{m\pi \over 2}), ~~
k^2_{ij} = {\pi^2 \over L^2} \left(i^2+j^2\right)
\eqno(10)
$$
where $l$ and $m$ are either 0 or 1 for cosine and sine waves,
$i$ and $j$ specify the order of Fourier modes, $N$ controls
the number of Fourier modes used, and $\kappa_{ijlm}$ are 
the surface density of fluctuations for the mode specified
by $i,j,l$, and $m$. Smaller and smaller
scale fluctuations are used when $N$ increases. We take
the side length of the box to be $4\arcsec$. This size is chosen
such that the periodic boundary condition does not influence 
the central region of the deflection potential.
Notice that the number of extra parameters is $4N^2$, therefore
for $N=1$, the number of degrees of freedom is already zero;
for $N\ge 2$, the system has more free parameters than 
constraints. The best fit parameters are
again found by minimizing a $\chi^2$ measure, which now utilizes
all the flux ratios (cf. Kormann et al. 1994b). To avoid excessively
large fluctuations, we include a term $\delta\chi^2 =
\sum_{i,j}\sum_{l,m}(\kappa_{ijlm}/0.05)^2$.
For the model presented in Table 1, we find
$\chi^2=82, 12, 6, 2$ for $N=0, 1, 2, 3$, while for $N\ge 4$, $\chi^2$
has essentially dropped to zero. For $N=0,1,2$, most $\chi^2$ are
from observed fluxes and positions.
It is interesting that for $N=1, 2$, when the number of parameters
is equal to or larger than the number of constraints, the $\chi^2$
remains high. This shows that large scale fluctuations are ineffective
in reducing the $\chi^2$, again highlighting the need for small-scale
structures. The fluctuation wavelengths corresponding
to $N=4, 5$ are $2\arcsec, 1\arcsecf 6$. Therefore
fluctuations with wavelength of $\la 1\arcsecf 5$
are sufficient to reproduce the observed flux ratios.
The resulting surface density distribution for $N=5$ is shown in
Fig.\ts 4 as the solid contours. It is obvious that the contours close
to the brighter images are somewhat twisted. In particular,
the surface density at the positions of 
image A and C are increased while 
that at the position of image B is decreased. The magnifications
are $15.3, -15.6, 8.1$ for A, B and C, respectively while that
for image D is essentially unchanged ($-$0.32). However,
notice in Fig.\ts 3, the outer density contours are 
moderately distorted. Such distortions can be avoided
by fluctuations with smaller wavelengths. Therefore smaller
scale fluctuations are preferred.

\xfigure{3}{The contours of constant surface density for
a singular isothermal ellipsoid with external shear and plane
wave fluctuations described by eq. (10), with $N=5$ and $L=4\arcsec$.
The contour levels
are 4, 2, 1, 0.5 and 0.25 of the critical surface density, respectively.
The thin lines show the corresponding contours for the underlying singular
isothermal ellipsoid.  The four crosses show the image positions and the central
dot shows the position of the primary lens galaxy, G1.
}{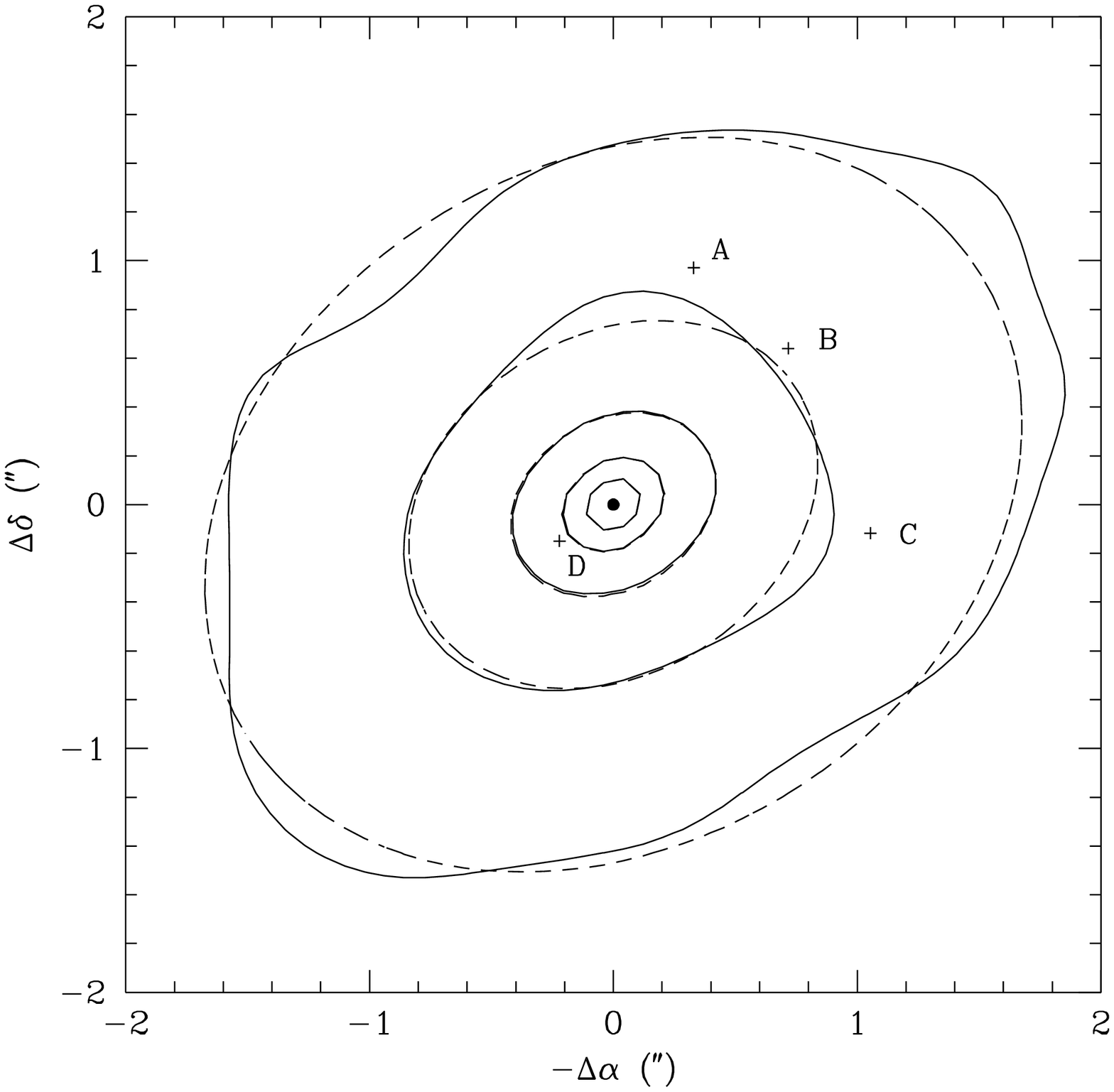}{10}

It should be noted that the model presented in Fig.\ts 3 is not at all
unique. The `best fit model' depends certainly on the macromodel from
which the $\chi^2$-minimization was started, and even for a fixed
starting model, there are many different models for which $\chi^2$ is
nearly zero for $N\ge 4$. Therefore, the model in Fig.\ts 3 should not
be viewed as a realistic lens model, but the exercise should
demonstrate the largest possible scale of substructure which can bring
the flux ratios within the observed range.

\sec{5 Practical modelling of lenses with substructure}

If we accept the view that the surface mass density of lens galaxies
is not as smooth as described by `simple' lens models, using flux
ratios for modelling may yield misleading conclusions. On the other
hand, discarding them altogether would result in a loss of
information; in particular, as long as the
perturbations are small, the magnification is only weakly affected
for images which are not highly magnified. Therefore, a statistical
treatment may be appropriate. In this section we propose a simple
method to perform lens modelling in the presence of small-scale
structure in the lens. 

Let $\hat p_\mu(\abs{\mu};\kappa_0,\abs{\gamma_0})$ be the
probability density for a magnification $\abs{\mu}$, given that the smooth
model has surface mass density $\kappa_0$ and (complex) 
shear $\gamma_0$. If the
unlensed flux of the source is $S_0$, the probability density for the
magnified flux $S=\abs{\mu} S_0$ becomes
$$
p_S(S;\kappa_0,\abs{\gamma_0})={1\over S_0}\,
\hat p_\mu\rund{{S\over S_0};\kappa_0,\abs{\gamma_0}}\; .
\eqno (11)
$$
Furthermore, let the observed flux be $S^\obs=x S$, where the
deviation of $x$ from unity accounts for measurement
uncertainties. Then, the likelihood to obtain a flux measurement
$S^\obs$ becomes
$$
\hat p_S(S^\obs;\kappa_0,\abs{\gamma_0})=\int_0^\infty 
{\d S\over S}\;p_S(S;\kappa_0,\abs{\gamma_0})\,p_x\rund{S^\obs\over S}\; ,
\eqno (12)
$$
where $p_x(x)$ is the probability density of the `measurement error' $x$.
Therefore, for $N$ images with
observed positions $\vc\theta_i^\obs$, positional uncertainty
$\sigma_i$, and observed fluxes $S^\obs_i$, the likelihood function that
can be maximized for obtaining model fits is
$$
\L=\prod_{i=1}^N {1\over \pi \sigma_i^2}
\exp\rund{-{\abs{\vc\theta_i-\vc\theta_i^\obs}
\over \sigma_i^2}}\,
\hat p_S\rund{S_i^\obs;\kappa_i,\gamma_i}\;,
\eqno (13)
$$
where $\vc\theta_i$ are the predicted image positions, and
$\kappa_i=\kappa_0(\vc\theta_i)$ and $\gamma_i=\abs{\gamma_0(\vc\theta_i)}$
are the surface mass density and shear at these positions.

For concreteness, we shall assume that the determinant $D\equiv
\det(\A) =(1-\kappa_0-\delta\kappa)^2-\abs{\gamma_0+\delta\gamma}^2$ is
distributed like a Gaussian with mean $\ave{D}$ and dispersion
$\sigma_D$. If, for example, the perturbation quantities
$\delta\kappa$ and $\delta\gamma$ were Gaussian,
$$
p(\delta\kappa,\delta\gamma)=
{1\over \sqrt{2}\pi^{3/2}\sigma_\kappa\sigma_\gamma^2}
\exp\rund{-{(\delta\kappa)^2\over
2\sigma_\kappa^2}-{\abs{\delta\gamma}^2\over \sigma_\gamma^2}}\; ,
\eqno (14)
$$
then $D$ follows very closely a Gaussian distribution, with
$$
\ave{D}=D_0+\sigma_\kappa^2-\sigma_\gamma^2\quad ;\quad
\sigma_D^2=4(1-\kappa_0)^2\sigma_\kappa^2+2\abs{\gamma_0}^2\sigma_\gamma^2 
+\sigma_\gamma^4+2\sigma_\kappa^4\; ,
\eqno (15)
$$
where $D_0=(1-\kappa_0)^2-\abs{\gamma_0}^2$ is the determinant for the
unperturbed lens. Since $\mu=1/D$, the probability density for
the magnification is $p_\mu(\mu)=p_D(1/\mu)/\mu^2$, and so the
probability density for the absolute value of $\mu$ becomes
$$
\hat p_\mu(\abs{\mu})={1\over\abs{\mu}^2}\eck{p_D\rund{1\over \abs{\mu}}
+p_D\rund{-1\over \abs{\mu}}}\; ;
\eqno (16)
$$
for notational simplicity we currently drop the arguments
$\kappa_0$ and $\gamma_0$. Assuming a Gaussian distribution for $x$, with
mean 1 and dispersion $\sigma_x$, we obtain from the foregoing
expressions that
$$
\hat p_S(S^\obs)={1\over 2\pi S_0\sigma_D\sigma_x\,a}
\rund{{\rm e}^{-b}+{\sqrt{\pi}\over 2}\eck{ g\rund{c_1+c_2\over
2\sqrt{a}} +g\rund{c_1-c_2\over 2\sqrt{a}}}}\;,
\eqno (17)
$$
and we introduced the abbreviations
$$\eqalign{
a&={1\over 2\sigma_D^2}+{(S^\obs/S_0)^2\over 2 \sigma_x^2} \quad ,
\quad 
b={\ave{D}^2\over 2 \sigma_D^2}+{1\over 2\sigma_x^2}\; ,\cr
c_1&={S^\obs\over S_0\,\sigma_x^2}\quad ,\quad
c_2={\ave{D}\over \sigma_D^2}\;, \cr
g(z)&=z\,{\rm e}^{z^2-b}\, {\rm erfc}(-z)\; , \cr }
\eqno (18)
$$
where ${\rm erfc}(z)$ is the complimentary error function. This
expression for $\hat p_S$ is easy to calculate, e.g., by using the
routine {\tt erfcc} of Press et al.\ts (1992).\fonote{A FORTRAN77
routine which calculates $\hat p_S(S^\obs)$ can be obtained from the
authors on request.} In Fig.\ts 4, we have
plotted the function $\hat p_S(S^\obs)$ for several combinations of
$\kappa_0$ and $\gamma_0$. As can be readily seen, the probability
distribution broadens considerably for larger values of the
magnification of the macromodel. Hence, the fluxes of weakly magnified
images provide strong 
constraints, whereas highly magnified images yield weaker constraints
on the model.

\xfigure{4}{The probability density $\hat p_S(S^\obs)$, according to
eq.\ts(17) of the text. The parameters of the macromodel was chosen to
be $\kappa_0=\gamma_0$, with $\kappa_0= 0.1$ (left-most curve), 0.2,
0.3, 0.4, 0.45, 0.48. The probability density for the perturbations
$\delta\kappa$ and $\delta\gamma$ have been assumed to follow a
Gaussian, with width $\sigma_\kappa=\sigma_\gamma=0.02$. Furthermore,
the measurement errors were characterized by a Gaussian distribution
in $x$ with width $\sigma_x=0.02$.  It is easily seen that images with
higher magnifications attain a much broader flux distribution than
those where the magnification is of order unity.
}{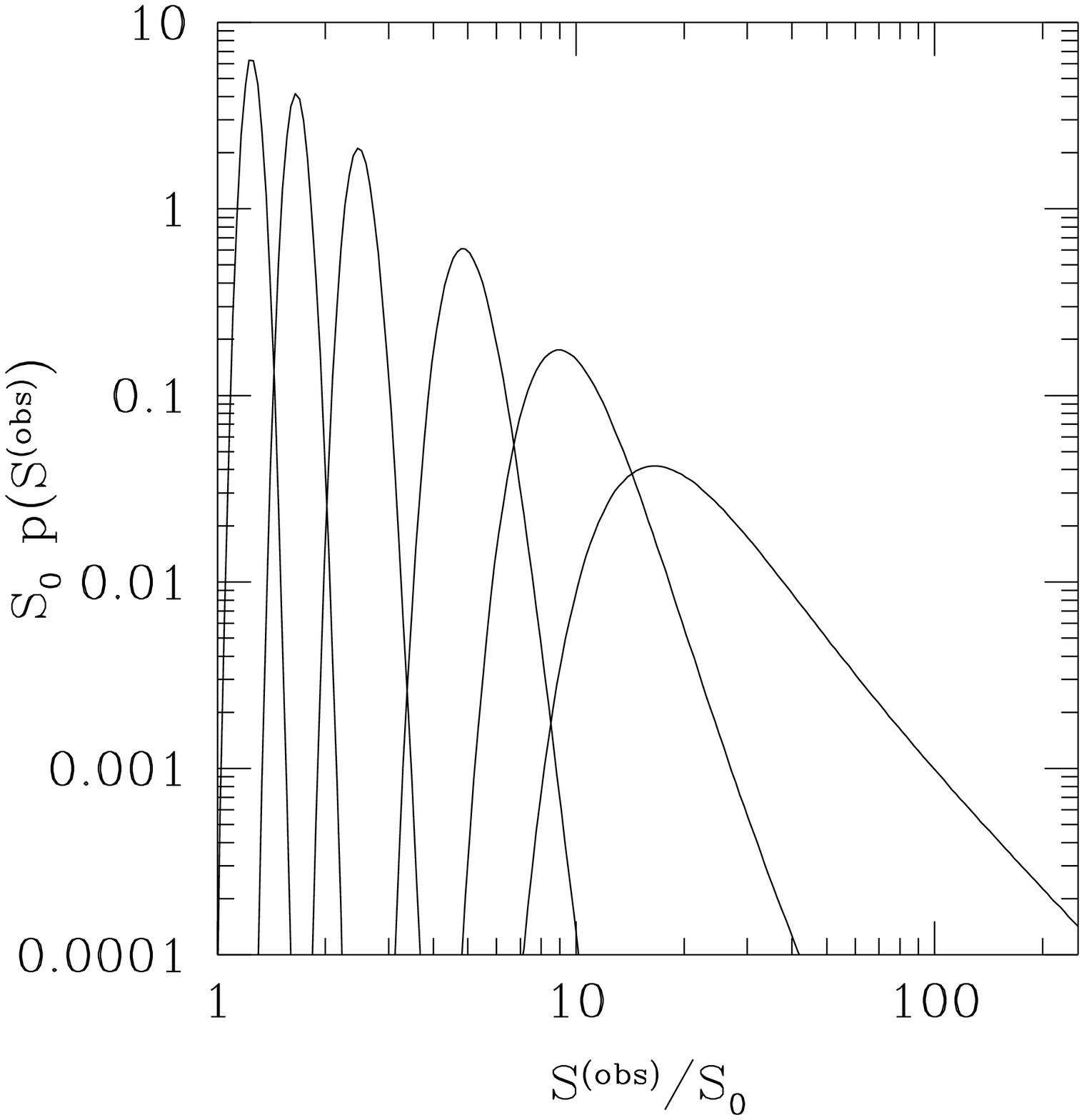}{10}

\vfil \eject
\sec{6 Discussion}

We have considered the possibility that the flux ratios in lens systems
may be affected by small scale structure in the lens galaxy. 
Although we discussed only the SIE model, we have verified that
it still seems difficult to fit the observed flux ratios using other
power-law models of the deflection potential. A general
analytic argument was presented why `simple' lens models are unable to
account for the radio flux ratios in B1422+231. We considered two
potential pictures for substructure, namely `globular clusters' with
$\sim 10^6M_\odot$ and smooth perturbations, modelled as plane density
waves. We have considered both pictures analytically, as well as through
numerical simulations. Owing to the large magnifications of images A,
B, C in 1422, only small perturbations (in mass) are needed to change
magnifications, and thus flux ratios appreciably.

The required perturbation is of the order of 
$\sim 1\%$ of the critical
surface density, or about $(40h M_\odot {\rm pc^{-2}})$ in physical
units. There are about 200 globular clusters
in the Galaxy with a total mass of $\approx 10^8 M_\odot$. These
globular clusters form a quasi-spherical distribution concentrated
toward the galactic center. The average surface density of these objects
within 5kpc is $\sim 1 M_\odot {\rm pc^{-2}}$. This is
$2.5\times 10^{-4} h^{-1}$ of the critical surface density, somewhat
too low to produce a significant change in magnification (cf. Fig.\ts 1). The
`globular cluster' picture should apply equally well to the hypothetical
massive black holes in galactic halos (e.g., Lacey \& Ostriker 1985).
They can produce much more visible deviations if just
a few percent of the halo mass are in these objects. Our second model
of substructure addresses smooth fluctuations. An obvious example of smooth
fluctuations is the spiral arms observed in disk galaxies. Smooth 
inhomogeneities are also quite naturally expected in
hierarchical structure formation models as a result of continuous
merging and accretion of sub-clumps.

An important question is
whether the added perturbation affects the time delay 
significantly: we found from models like those shown in Figs.\ts 2 and 3 that
the perturbation changes the 
time delay at most by a few percent. For smaller-scale perturbations than
the example shown in Fig.\ts 3, the
time delays will be affected even less.
Therefore we conclude that substructure
can have only minor
effects on the time delay -- gravitational lensing
is still potentially a golden tool for determining the Hubble constant.

We thank Matthias Bartelmann, Simon White and Hongsheng Zhao
for helpful discussions and comments on the paper.
\SFB

\def\ref#1{\vskip1pt\noindent\hangindent=40pt\hangafter=1 {#1}\par}
\sec{References}
\ref{Chang, K., Refsdal, S., 1979, Nature, 282, 561}
\ref{Courbin, F., et al., 1997, astro-ph/9705093}
\ref{Falco, E.E., Leh{\'a}r, J., Perley, R.A., Wambsganss, J.,
Gorenstein, M.V., 1996, AJ, 112, 897}
\ref{Falco, E.E., Leh\'ar, J., Shapiro, I.I., 1996, astro-ph/9612182}
\ref{Falco, E.E., Shapiro, I.I., Moustakas, L.A., Davis, M., 1997,
astro-ph/9702152}
\ref{Hogg, D.W., Blandford, R.D., 1994, MNRAS, 268, 889}
\ref{Impey, C.D., Foltz, C.B., Petry, C.E., Browne, I.W.A.,
Patnaik, A.R., 1996, ApJ, 462, L53}
\ref{Keeton II, C.R., Kochanek, C.S., 1996, astro-ph/9611216}
\ref{Keeton II, C.R., Kochanek, C.S., Seljak, U., 1997, ApJ, 482, 604}
\ref{Kent, S.M., Falco, E.E., 1988, AJ, 96, 1570}
\ref{Kochanek, C.S., 1991, \apj, 373, 354}
\ref{Kormann, R., Schneider, P., Bartelmann, M., 1994a,
A\&A, 284, 285}
\ref{Kormann, R., Schneider, P., Bartelmann, M., 1994b,
A\&A, 286, 357}
\ref{Kundic, T., Hogg, D.W., Blandford, R.D., Cohen, J.G., Lubin,
L.M., Larkin, J.E., 1997, astro-ph/9706169}
\ref{Lacey, C. G., Ostriker, J. P., 1985, 290, 154}
\ref{Mao, S., 1992, ApJ, 389,  63}
\ref{Nair, S., Garrett, M.A., 1997, MNRAS 284, 58.}
\ref{Patnaik, A.R. et al., 1992, \mnras, 259, 1$_{\rm P}$ (P92)}
\ref{Patnaik, A.R., Procas, R.W., 1997, preprint}
\ref{Refsdal, S., 1964, MNRAS, 128, 307}
\ref{Rix, H.-W., Schneider, D.P., Bahcall, J.N., 1992, AJ, 104, 959}
\ref{Schneider, P., Ehlers, J., Falco, E. E., 1992, 
                  {\it Gravitational Lenses} (Springer-Verlag: New York)}
\ref{Schneider, P., Weiss, A., 1992, A\&A, 260, 1}
\ref{Tonry, J. L. 1997, astro-ph/9706199}
\ref{Wallington, S., Kochanek, C.S., Narayan, R., 1996, ApJ, 465, 64} 
\ref{Wambsganss, J., 1990, report MPA 550, Garching}
\ref{Wambsganss, J., Paczy\'nski, B., 1992, ApJ, 397, L1}
\ref{Wambsganss, J., Paczy\'nski, B., 1994, \aj, 108, 1156}
\ref{Witt, H.J., Mao, S., Schechter, P., 1995, ApJ, 443, 18}
\ref{Zwicky, F., 1937, Phys. Rev., 51, 290}

\vfill\eject\end